\documentclass[aps,prl,10pt,twocolumn,superscriptaddress]{revtex4-1}

\usepackage{amsmath, amssymb}    					
\usepackage{graphicx}   							
\usepackage{xcolor}     							
\usepackage{hyperref}   							
\hypersetup{colorlinks=true,allcolors=blue}			
\usepackage{textcomp}								

\renewcommand{\t}[1]{\textrm{#1}}

\renewcommand{\Im}{\textrm{Im}}
\renewcommand{\Re}{\textrm{Re}}

\begin{document}

\title{Many-body correlations brought to light in absorption spectra of diluted magnetic semiconductors}
\author{F. Ungar}
\affiliation{Theoretische Physik III, Universit\"at Bayreuth, 95440 Bayreuth, Germany}
\author{M. Cygorek}
\affiliation{Department of Physics, University of Ottawa, Ottawa, Ontario, Canada K1N 6N5}
\author{V. M. Axt}
\affiliation{Theoretische Physik III, Universit\"at Bayreuth, 95440 Bayreuth, Germany}

\begin{abstract}

Diluted magnetic semiconductors are materials well known to exhibit strong correlations which typically manifest in carrier-mediated magnetic ordering.
In this Rapid Communication, we show that the interaction between excitons and magnetic impurities in these materials is even strong enough to cause a significant deviation
from the bare exciton picture in linear absorption spectra of quantum well nanostructures.
It is found that exciton-impurity correlations induce a characteristic fingerprint in the form of an additional side structure close to the exciton resonance 
in combination with a shift of the main exciton line of up to a few meV.
We trace back these structures to the form of the self-energy and demonstrate that reliable values of the average correlation energy per exciton can be extracted directly 
from the spectra.
Since the only requirement for our findings is sufficiently strong correlations, the results can be generalized to other strongly correlated systems.

\end{abstract}

\maketitle


Many-body correlations are an important and extensively studied phenomenon in many areas of physics \cite{Karaiskaj_Two-Quantum_2010, Basov_Electrodynamics-of_2011, 
Chemla_Many-body_2001, Shumway_Correlation-versus_2001, Hybertsen_Electron-correlation_1986, DiMarco_Electron-correlations_2013, Goerg_Enhancement-and_2018,
Ma_Uncovering-many-body_2014, Mentink_Ultrafast-and_2015, Walther_Giant-optical_2018, Cronenwett_A-Tunable_1998, Yazyev_Mangetic-Correlations_2008, 
Qazilbash_Electronic-correlations_2009, Ossiander_Attosecond-correlation_2016, Bonitz_Complex-plasmas_2010},
the impact of which is typically investigated using nonlinear response such as four-wave mixing \cite{Smith_Extraction-of_2010, Oestreich_Exciton-Exciton_1995, 
Kner_Coherence-of_1998, Axt_How-Correlated_2000, Bolton_Demonstration-of_2000, Tahara_Non-Markovian_2011}.
An interesting subclass of materials that are known for strong correlation effects are diluted magnetic semiconductors (DMSs), i.e., II-VI or III-V semiconductor alloys 
with a small percentage of magnetic dopants, typically manganese \cite{Bouzerar_Unified-picture_2010, Dietl_Dilute-ferromagnetic_2014, Ohno_GaMnAs-A_1996, 
Kossut_Introduction-to_2010, Furdyna_Diluted-magnetic_1988}.
Most notably in these materials, correlations have been found to cause carrier-mediated ferromagnetic ordering \cite{Ohno_Making-Nonmagnetic_1998}, a topic which is still
actively investigated \cite{DiMarco_Electron-correlations_2013}.
DMSs are also known for possible applications in the field of spintronics \cite{Awschalom_Challenges-for_2007, Dietl_A-ten_2010, Ohno_A-window_2010, 
Zutic_Spintronics-Fundamentals_2004, Joshi_Spintronics_2016}, either as a spin aligner \cite{Fiederling_Injection-and_1999} or in terms of data storage applications
\cite{Chappert_The-emergence_2007}.

The physics in DMSs is typically dominated by the strong exchange interaction between carriers and magnetic dopants which is usually modeled by a 
Kondo-type Hamiltonian for electrons and holes with well-established coupling constants \cite{Kossut_Introduction-to_2010}.
Since typical experiments on DMS nanostructures are performed close to the exciton resonance \cite{Camilleri_Electron-and_2001, 
Roennburg_Motional-Narrowing_2006, Crooker_Optical-spin_1997, BenCheikh_Electron-spin_2013, Akimoto_Coherent-spin_1997, Murayama_Spin-dynamics_2006, 
Smits_Excitonic-enhancement_2004, Sakurai_Ultrafast-exciton_2004}, it is evident that effects due to the Coulomb interaction cannot be neglected.
In addition, it was already pointed out in the literature that a mean-field treatment of the electron-impurity exchange interaction is often insufficient
for an accurate description of ultrafast spin dynamics \cite{Thurn_Non-Markovian_2013, Cygorek_Influence-of_2017} as well as order parameters 
\cite{Sato_First-principles_2010, DiMarco_Electron-correlations_2013}.

In this Rapid Communication we explore the impact of exciton-impurity correlations on linear absorption spectra of DMS nanostructures close to the exciton resonance.
Considering that carrier-impurity correlations typically manifest themselves in the dynamical properties of DMSs such as spin overshoots on picosecond time scales 
\cite{Thurn_Non-Markovian_2013, Ungar_Quantum-kinetic_2017} or modifications of spin-transfer rates \cite{Ungar_Trend-reversal_2018}, it is often challenging to 
pinpoint these features in experiments since they can be extremely dependent on the particular sample.
This is, e.g., due to the variation of the degree of band mixing between light and heavy holes which may cause hole spin relaxation times to vary between a few 
picoseconds \cite{Crooker_Optical-spin_1997} up to longer than the exciton lifetime \cite{Camilleri_Electron-and_2001}, which makes it hard to accurately 
predict experimental observations.
Furthermore, it is typically not possible to quantify the amount of correlation energy that has built up in the sample since correlations
often merely manifest in bandgap renormalizations or influence spectral widths and are thus hard to isolate \cite{Hanbicki_Measurement-of_2015, Ugeda_Giant-bandgap_2014, 
Kwong_Self-consistent_2009, Piermarocchi_Role-of_2001, Jahnke_Influence-of_2000}.
It is therefore of particular interest to be able to identify many-body correlation effects in an experimentally well accessible quantity such as the linear absorption 
spectrum.
Indeed, we show that correlation effects are visible in the spectrum of DMSs and lead to a side structure close to the main exciton peak accompanied by a shift of the bare exciton
resonance.
Our calculations reveal that the absorption spectrum provides direct access to the average correlation energy per exciton, which is otherwise hard to measure,
and thus allows for an estimation of the importance of many-body effects without having to resort to high excitation powers.


We consider a narrow DMS quantum well that is excited close to the $1s$ exciton resonance below the band gap.
Furthermore, we focus on the widely studied class of II-VI DMSs where the impurity ions are isoelectronic so that no excess charge
carriers are present in the system.
Accounting only for the energetically lowest confinement state with respect to the growth direction, which is justified for sufficiently narrow quantum wells 
\cite{Astakhov_Binding-energy_2002}, we obtain the wave function and the binding energy of the exciton ground state by solving the corresponding Schr{\" o}dinger
equation for electrons and holes bound by the Coulomb interaction.
The interactions that typically dominate the physics of DMSs are the $s$-$d$ and $p$-$d$ exchange mechanisms between $s$-type conduction band electrons or 
$p$-type valence band holes and the localized $d$-shell electrons of the impurity ions, respectively \cite{Dietl_Dilute-ferromagnetic_2014, Kossut_Introduction-to_2010, 
Furdyna_Diluted-magnetic_1988, Sato_First-principles_2010}.
This interaction describes the scattering of electrons and holes accompanied by a simultaneous spin-flip event of the respective carriers and the magnetic dopants.
We also account for a purely nonmagnetic type of interaction between carriers and impurities that arises, e.g., due to the band-gap mismatch when doping atoms 
are incorporated into the host lattice.
Finally, we fully take into account the optical excitation in the dipole approximation.
The explicit expressions for all parts of the Hamiltonian can be found in the Supplemental Material \cite{supplement}.

Based on the model described above, a quantum kinetic theory for the exciton spin dynamics which explicitly includes the optical coherence has been developed 
in Ref.~\cite{Ungar_Quantum-kinetic_2017}.
Using a formulation in terms of excitonic density matrices together with a correlation expansion to treat higher-order expectation values between exciton and impurity
operators, exciton-impurity correlations can be kept explicitly as dynamical variables.
A closed set of equations of motion is obtained by invoking the dynamics-controlled truncation (DCT) \cite{Axt_A-dynamics_1994, Axt_Nonlinear-optics_1998} so that 
variables up to second order in the laser field are taken into account.
Here, we use this theory to calculate the linear absorption of a DMS nanostructure beyond the single-particle level.

Since optical spectra may also be affected by phonons, we extend the theory of Ref.~\onlinecite{Ungar_Quantum-kinetic_2017} to also account for the influence of 
longitudinal acoustic phonons via deformation potential coupling, which typically dominates the linewidth in semiconductors for temperatures below $80\,$K 
\cite{Rudin_temperature-dependent_1990}.
We limit the description to bulk phonons which is justified because of the rather weak dependence of the lattice constant on the impurity content within the considered 
doping range \cite{Furdyna_Diluted-magnetic_1988}.
To obtain the absorption, we set up the equation of motion for the excitonic interband coherence $y = \langle \hat{Y}_{1s} \rangle$, where $\hat{Y}_{1s}$
denotes the annihilation operator of an exciton in the $1s$ ground state with a vanishing center-of-mass wave vector.
The appearing source terms are given by impurity- and phonon-assisted variables for which separate equations of motion have to be set up.
A formal integration of the latter yields an integro-differential equation which can be solved in Fourier space to obtain the susceptibility
\cite{Axt_Nonlinear-optics_1998, Thraenhardt_Quantum-theory_2000, Kira_Many-body_2006}.
The details of the derivation as well as the resulting equations can be found in the Supplemental Material \cite{supplement}.

\begin{figure}[t!]
\centering
\includegraphics{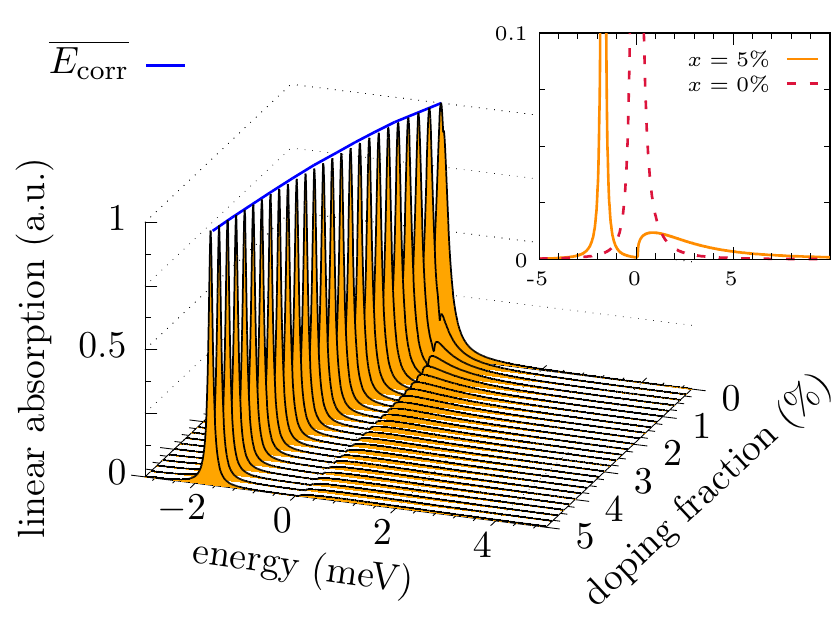}
\caption{Linear absorption spectra of a $15\,$nm wide Zn$_{1-x}$Mn$_{x}$Se quantum well at $30\,$K for various doping fractions $x$.
		 The energy scale is chosen such that $E = 0$ coincides with the maximum of the $1s$ exciton resonance at $x = 0$ and the 
		 spectra are normalized with respect to each $1s$ absorption peak.
		 The average correlation energy per exciton $\overline{E_\t{corr}}$ obtained numerically from the full quantum kinetic model is indicated with respect
		 to the maximum of each side structure.
		 The inset shows a magnified view of the absorption spectra without doping and with an impurity content $x = 5\%$.}
\label{fig:xMnvar}
\end{figure}

For the numerical simulations we focus on Zn$_{1-x}$Mn$_{x}$Se quantum wells of varying widths and doping fractions $x$ at a temperature of $30\,$K.
Since typical exciton lifetimes vary between several $10\,$ps to $100\,$ps \cite{Poltavtsev_Damping-of_2017, Chen_Spin-dynamics_2003, Runge_Relaxation-kinetics_1995,
Kalt_Optical-and_1992}
we also include a radiative decay rate of $0.1\,$ps$^{-1}$ which affects excitons close to the bottom of the exciton parabola \cite{Siantidis_Dynamics-of_2001, 
Thraenhardt_Quantum-theory_2000}.
Otherwise, standard parameters for a ZnSe-based semiconductor nanostructure and the necessary coupling constants are used \cite{Furdyna_Diluted-magnetic_1988, 
Astakhov_Binding-energy_2002, Cardona_Acoustic-deformation_1987, Strzalkowski_Dielectric-constant_1976}.

Figure~\ref{fig:xMnvar} shows the calculated linear absorption spectra of a $15\,$nm wide quantum well for doping concentrations between $0\%$ (pure nonmagnetic ZnSe) 
and $5\%$.
As expected, in the case of undoped ZnSe, a single peak appears at the $1s$ exciton resonance which is broadened due to radiative decay and slightly asymmetric due to 
the phonon influence (cf. the inset of Fig.~\ref{fig:xMnvar}).
Choosing the origin of the energy scale to coincide with the maximum of that peak, the exciton line becomes redshifted upon an increase of the doping fraction
and lies at approximately $-1.8\,$meV at an impurity content of $5\%$.
In addition, a second feature appears in the spectrum that is completely absent for an undoped quantum well and splits off from the main exciton resonance with
increasing doping fraction.
It becomes gradually smeared out for larger doping fractions and shows an exponentially decaying tail on the high-energy side.

The average correlation energy per exciton can be defined as
\begin{align}
\label{eq:Ecorr}
\overline{E_\t{corr}} = \frac{1}{T} \int_{0}^{T}dt\, \frac{\langle H \rangle(t) - \langle H \rangle^\t{mf}(t)}{n_\t{X}(t)},
\end{align}
where $n_\t{X}(t)$ denotes the number of excitons, $\langle H \rangle(t)$ is the expectation value of the complete Hamiltonian from which the mean-field contribution
$\langle H \rangle^\t{mf}(t)$ is subtracted, and $T$ is the averaging time.
As shown in the Supplemental Material, $\overline{E_\t{corr}}$ becomes almost independent on $T$ after a short initial period where the correlations build up after the
pulsed excitation \cite{supplement}.
A comparison of $\overline{E_\t{corr}}$ calculated using the full quantum kinetic model (cf. blue curve in Fig.~\ref{fig:xMnvar}) with the position 
of the exciton peak relative to the maximum of the second structure in the linear absorption spectra reveals an excellent agreement.
This shows that the observed shift of the exciton line and the second structure in the spectrum are indeed due to the influence of exciton-impurity correlations, 
an effect that cannot be obtained on the mean-field level.
In accordance with energy conservation, the build up of a negative correlation energy is accompanied by an increase in the average kinetic energy of excitons which
manifests in a redistribution of exciton momenta away from the optically active state with $K \approx 0$.
A more intuitive explanation of the correlation energy can be given in terms of energy eigenstates:
Since the exchange interaction as well as the nonmagnetic impurity scattering couples the exciton and the impurity system, the bare exciton states are no longer 
the proper eigenstates of the many-body system, resulting in the observed modification of the linear absorption.

Numerical calculations confirm that the two structures remain visible even at liquid nitrogen temperatures of up to $77\,$K.
This is due to the fact that the phonon-induced broadening of the exciton line is rather small for acoustic phonons, a result which is in line with experimental data
as well as theoretical calculations where an increase in the half width at half maximum (HWHM) linewidth of only a few {\textmu eV} per K has been found 
\cite{Rudin_Effects-of_2002, Gammon_Phonon-broadening_1995}.
Thus, as long as temperatures below the longitudinal optic (LO)-phonon threshold are considered, phonons have a negligible influence on the exciton line regarding the phenomena
discussed here.
It should be noted that acoustic phonons cause the exciton line to become increasingly asymmetric with rising temperature since more and more states become
accessible on the exciton parabola \cite{Rudin_Effects-of_2002}.
Phonons also provide a small contribution to the redshift of the exciton line on the order of $10\,$\textmu eV at $30\,$K, which is clearly negligible compared 
with the shift due to impurity scattering.

\begin{figure}[t!]
\centering
\includegraphics{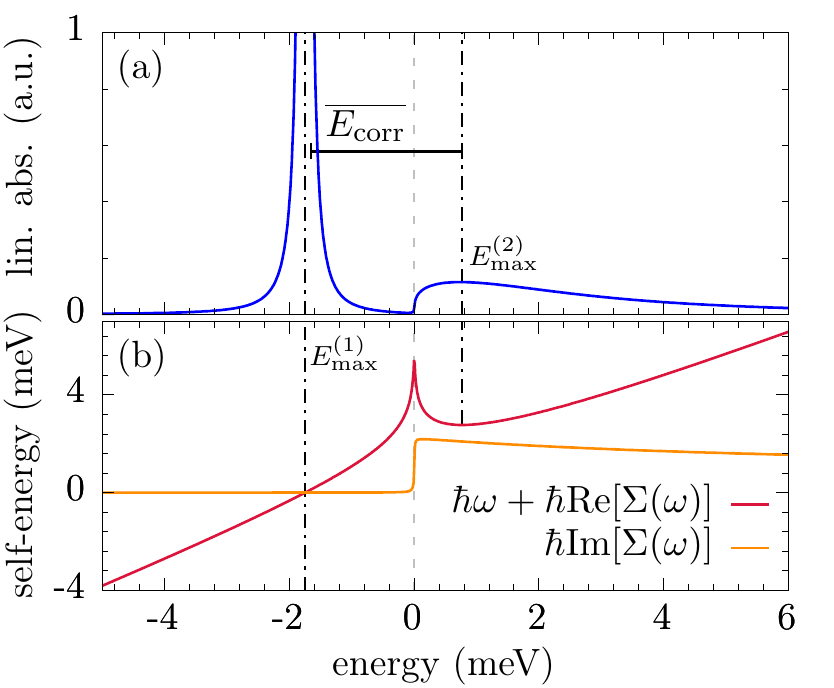}
\caption{Explanation of the absorption lineshape. We compare (a) the linear absorption spectrum obtained from Eq.~\eqref{eq:alpha} with
		 (b) the self-energy $\Sigma(\omega)$ for a $15\,$nm wide Zn$_{0.95}$Mn$_{0.05}$Se quantum well.
		 The self-energy can be used to explain the position of the main exciton resonance ($E_\t{max}^{(1)}$) as well as the maximum of the second structure
		 in the spectrum ($E_\t{max}^{(2)}$).
		 The average correlation energy per exciton $\overline{E_\t{corr}}$ calculated numerically from the full model is indicated by the length of the horizontal bar
		 with respect to $E_\t{max}^{(2)}$.}
\label{fig:lineshape}
\end{figure}

In order to obtain a better understanding of the observed structure of the linear absorption spectrum, it is thus justified to focus on a system without
phonon influence.
Then, the linear absorption is found to be
\begin{align}
\label{eq:alpha}
\alpha(\omega) \sim \frac{\Gamma_0 + \Im[\Sigma(\omega)]}{\big(\omega + \Re[\Sigma(\omega)]\big)^2 + \big(\Gamma_0 + \Im[\Sigma(\omega)]\big)^2}
\end{align}
with a complex self-energy $\hbar\Sigma(\omega)$ and the radiative decay rate $\Gamma_0$ \cite{Rudin_Effects-of_2002, Wachter_Excitation-induced_2002, 
Mannarini_Near-field_2006, Kwong_Self-consistent_2009}.
For an explicit expression of the self-energy the reader is referred to the Supplemental Material \cite{supplement}.
A comparison of $\alpha(\omega)$ with the structure of the self-energy in Fig.~\ref{fig:lineshape} indeed reveals that the physics behind the linear absorption spectra 
can be well understood by Eq.~\eqref{eq:alpha}.
Since $\Im[\Sigma(\omega)]$ is strictly positive it follows that the strongest resonance is determined by the condition $\omega + \Re[\Sigma(\omega)] = 0$.
Thus, the real part of the self-energy causes the shift of the exciton line whose width is determined by the rate $\Gamma_0$ since $\Im[\Sigma(\omega)]=0$
for $\omega < 0$ (cf. Fig.~\ref{fig:lineshape}).
In contrast, the second structure in the spectrum is exclusively due to the imaginary part of the self-energy which becomes finite for $\omega \gtrsim 0$ and slowly decreases 
for larger energies.
Figure~\ref{fig:lineshape} confirms that the onset of the second structure indeed corresponds to the rise of $\Im[\Sigma(\omega)]$ and the maximum of the structure
occurs at the minimum of $\omega + \Re[\Sigma(\omega)]$.
Note that, if the spectra for finite doping fractions would consist of just a single discrete line shifted from its position at $x=0\%$ by $\overline{E_\t{corr}}$, 
the interpretation of the spectral shift would remain valid but it would be challenging to extract the correlation energy per exciton since it would be necessary to compare 
spectra for different doping fractions in order to determine the shift.
Such experiments can easily become inconclusive since it is hard to change the doping fraction without changing other sample characteristics such as, e.g., internal
stress properties.

From a physical perspective, the elastic scattering at the impurity ions couples the optically dipole-allowed exciton states with vanishing wave vector to states
on the exciton parabola with $K > 0$, resulting in the excitation of a many-body state that has both contributions.
Such a mixing between $K = 0$ and $K \neq 0$ states due to disorder in the sample and its impact on the absorption spectrum has already been discussed in the literature
in the context of semiconductor quantum wells with magnetic barriers on a mean-field level, i.e., without accounting for exciton-impurity correlations \cite{Sugarov_Magnetic-field_2001,
Komarov_Magnetic-field_2006}.
There, disorder has been found to lead to different shapes of the exciton line when comparing the $\sigma^+$ with the $\sigma^-$ component of the exciton transition.
A study of excitons in rough quantum wells based on random potentials \cite{Zimmermann_Theory-of_1992} has also found a similar impact on the optical density, namely a 
shift of the exciton peak towards lower energies combined with an asymmetry towards the high-energy side.
However, the fluctuations of the random potential considered in that work were not strong enough to observe two distinct structures.
In contrast, the values for the exchange interaction in DMSs on the order of a few $10\,$meV are well established in the literature \cite{Dietl_Dilute-ferromagnetic_2014,
Kossut_Introduction-to_2010, Furdyna_Diluted-magnetic_1988} and are sufficiently large to cause a drastically different spectrum.

It should be noted that the scattering with phonons can also be cast into a form similar to Eq.~\eqref{eq:alpha} so that a self-energy can be identified 
\cite{Rudin_Effects-of_2002, Mannarini_Near-field_2006}.
However, there are some fundamental differences compared with the exciton-impurity scattering:
First, phonon scattering is an inelastic process, causing the phonon dispersion to appear in the self-energy.
Second, phonons introduce two contributions that can be interpreted as phonon absorption and emission, respectively.
The phonon self-energy leads primarily to an asymmetry of the exciton line at elevated temperatures and, as mentioned above, only causes a marginal shift of the exciton 
line due to the much smaller carrier-phonon coupling compared with the exchange interaction in DMSs.
Thus, in a theory accounting only for phonons, the exciton peak as well as the onset of a finite imaginary part of the self-energy do not appear as
separate features in the spectra.

\begin{figure}[t!]
\centering
\includegraphics{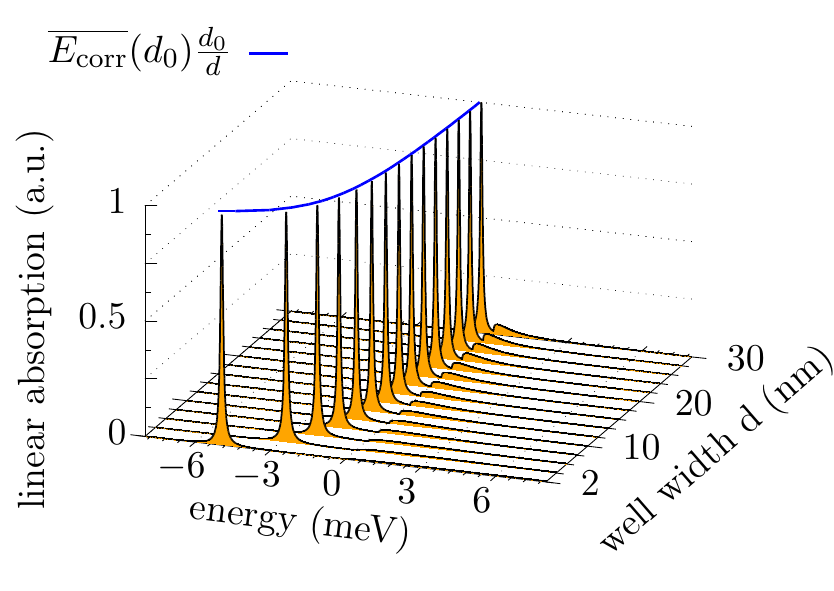}
\caption{Linear absorption spectra of a Zn$_{0.98}$Mn$_{0.02}$Se quantum well at $30\,$K for various quantum well widths $d$.
		 The energy scale is the same as in Fig.~\ref{fig:xMnvar} and all spectra a normalized with respect to each $1s$ absorption peak.
		 The blue line shows the inverse dependence of the correlation energy on the well width with respect to the reference point $d_0 = 15\,$nm.}
\label{fig:dvar}
\end{figure}

A study of the dependence of the spectra on the width of the DMS nanostructure, as depicted in Fig.~\ref{fig:dvar} for a Zn$_{0.98}$Mn$_{0.02}$Se 
quantum well at $30\,$K, reveals that the correlation-induced side structure becomes significantly broadened for smaller well widths and thus appears
more pronounced and localized for moderate widths of about $10\,$nm and above.
It is also found that the side structure starts to merge with the main exciton peak for larger well widths so that only a single line is to be expected in the
bulk limit.
The blueshift of the exciton line that occurs when the well width becomes larger corresponds to a decrease of the average correlation energy per exciton and shows
an inverse dependence on the quantum well width, which suggests that correlations are enhanced in smaller nanostructures and become less significant in bulk.
The scaling of the average correlation energy per exciton with $\frac{1}{d}$ follows directly from the prefactors of the complete expression for the correlation energy
\cite{supplement}.

\begin{figure}[t!]
\centering
\includegraphics{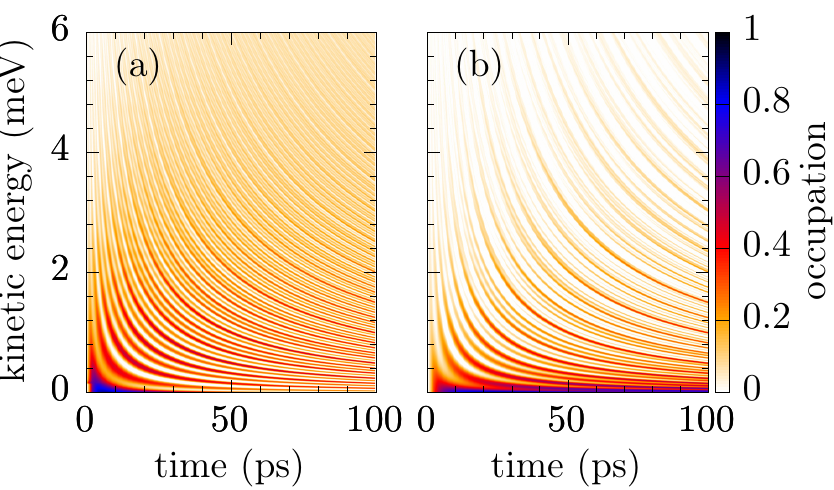}
\caption{Time evolution of the energy-resolved occupation of the $1s$ exciton parabola after optical excitation using a pulse with 
		 $0.1\,$ps FWHM for a $10\,$nm wide Zn$_{0.99}$Mn$_{0.01}$Se quantum well.
		 We compare (a) a calculation with phonons at a temperature of $80\,$K with (b) the phonon-free case.}
\label{fig:occupation}
\end{figure}

The fact that the formation of carrier-impurity correlations is accompanied by an occupation of states on the exciton parabola with a finite center-of-mass wave
number $K$ is confirmed by Fig.~\ref{fig:occupation}, where the time- and energy-resolved occupation of the exciton ground state is shown.
Without phonons, the only mechanism that can change the exciton wave vector is the elastic scattering at the impurities.
However, in the typically employed Markov approximation, this scattering is energy conserving so that an initial exciton occupation at $K \approx 0$ would
always remain at the bottom of the exciton parabola.
Instead Fig.~\ref{fig:occupation}(b) reveals that, even in the absence of phonons, a significant scattering towards higher center-of-mass energies takes place,
which can be associated with many-body correlations between the excitons and impurities that remain finite even for long times and cause a deviation from an effective
single-particle picture.
Including phonons further enhances this redistribution, especially for higher temperatures [cf. Fig.~\ref{fig:occupation}(a)].
This effect should be observable in experiments by, e.g., LO-phonon-assisted photoluminescence, which has already been successfully demonstrated for undoped ZnSe 
nanostructures \cite{Zhao_Coherence-length_2002, Zhao_Energy-relaxation_2002, Umlauff_Direct-observation_1998}.


All in all, we expect that the fingerprint of exciton-impurity correlations in absorption spectra of DMSs can be experimentally resolved provided that the exciton linewidth 
stays below a few meV.
We have shown that a correlation-induced side structure appears in the spectrum on the high-energy side of the $1s$ exciton line which is particularly pronounced for
quantum wells in the $10\,$nm range and impurity concentrations of a few percent.
Furthermore, the shift of the exciton line with respect to this side structure yields a reliable value for the average correlation energy per exciton from a single spectrum 
without having to compare different samples.
Since the only requirement for our findings is sufficiently strong correlations, our results can be generalized to many other correlated systems.


Financial support of the Deutsche Forschungsgemeinschaft (DFG) through Grant No. AX17/10-1 is gratefully acknowledged.

\bibliography{references}
\end{document}